\documentclass[12pt]{article}
\usepackage{epsfig}

\newcommand{\bea}{\begin{eqnarray}}
\newcommand{\eea}{\end{eqnarray}}
\newcommand{\be}{\begin{equation}}
\newcommand{\ee}{\end{equation}}
\newcommand{\MSbar}{\overline{\rm MS}}
\newcommand{\as}{\alpha_s}
\newcommand{\asMZ}{\alpha_s(M^2_Z)}
\newcommand{\ar}{a_s}
\newcommand{\dd}{\mathrm{d}}

\title{Effect of the
analytic and ``frozen'' coupling constants in QCD
up to NNLO from DIS data
\author{A.V.~Kotikov,  V.G.~Krivokhizhin,
and B.G.~Shaikhatdenov\\
Joint Institute for Nuclear Research, Russia}}

\begin{document}
\maketitle
\abstract{ We give a short review of our recent analysis \cite{KKB} of the
deep inelastic scattering data (provided by BCDMS, SLAC, NMC) on $F_2$ structure function in the non-singlet approximation with up to next-to-next-to-leading-order accuracy and analytic and frozen modifications of the strong coupling constant 
featuring no unphysical singularity (the Landau pole).
Improvement of agreement between theory and
experiment, with respect to the case of the standard perturbative
definition of $\as$ considered recently in \cite{BKKP}, was observed.
} \\

\section{ Introduction }

At present an accuracy of the data on the deep inelastic scattering (DIS)
structure functions (SFs) makes it possible to study separately the 
$Q^2$-dependence of logarithmic QCD-inspired corrections and those of power-like
(non-perturbative) nature (see for instance~\cite{Beneke} and references 
therein). A part of the power corrections may be responsible for the deviation
of the ``physical'' strong coupling constant from the standard $\MSbar$ one.
Similar effect was observed recently \cite{CIKK09}
in the region of small $x$. In a sense, the study done in \cite{KKB} is the 
extension of the 
analysis \cite{CIKK09} to the region of the intermediate and large $x$ values.

In the present paper we show the results of our recent analyse \cite{KKB} of
of DIS SF $F_2(x,Q^2)$ with SLAC, NMC and
BCDMS 
experimental
data involved~\cite{SLAC1}--\cite{BCDMS1} at NNLO of massless perturbative QCD.
This has become possible thanks to the results on both the $\alpha_s^3(Q^2)$ 
corrections to the splitting functions (the anomalous dimensions of Wilson 
operators)~\cite{MVV2004} and the corresponding expressions of the complete 
three-loop coefficient functions for the structure functions 
$F_2$ and $F_L$~\cite{MVV2005}.
\footnote{For the odd $n$ values, the corresponding coefficients 
can be obtained by using the analytic
continuation~\cite{KaKo,KoVe}.
}

As in our previous paper the function $F_2(x,Q^2)$ is represented as a sum of the leading twist $F_2^{pQCD}(x,Q^2)$ and the twist four terms:
\be
F_2(x,Q^2)=F_2^{pQCD}(x,Q^2)\left(1+\frac{\tilde h_4(x)}{Q^2}\right)\,.
\label{1.1}
\ee

As is known there are at least two ways to perform QCD analysis over DIS data: the first one (see e.g.~\cite{ViMi,fits}) deals
with Dokshitzer-Gribov-Lipatov-Altarelli-Parisi (DGLAP) integro-differential equations~\cite{DGLAP} and let the data be examined directly, whereas the second one involves the SF moments and permits performing an analysis in analytic form as opposed to the former option.
In this work we take on the way in-between these two latter, i.e. analysis is carried out over the moments of SF $F_2^{k}(x,Q^2)$ defined as follows
\be
M_n^{k}(Q^2)=\int_0^1 x^{n-2}\,F_2^{k}(x,Q^2)\,dx
~~~~ (k=\mbox{pQCD, twist2} \ldots)
\label{1.a}
\ee
and then reconstruct SF for each $Q^2$ by using Jacobi polynomial expansion 
method \cite{Barker,Kri} (for further details see~\cite{KK2001,KK2009}).
The theoretical input can be found
in the papers~\cite{KK2009,Kotikov:2007ua}.

\section{Infrared modifications of the strong coupling constant}

Here, we investigate the potential of modifying the
strong-coupling constant in the infrared region with the purpose
of illuminating the problem related to the Landau singularity in
QCD.

Specifically, we consider two modifications, which effectively
increase the argument of the strong-coupling constant at small
$Q^2$ values, in accordance with~\cite{large}.

In the first case, which is more phenomenological, we introduce freezing
of the strong-coupling constant by changing its argument as
$Q^2 \to Q^2 + M^2_{\rho}$, where $M_{\rho}$ is the rho-meson mass~\cite{frozen} (see also~\cite{Zotov,ShiTer} and references therein).
Thus, in the formulas of the previous section
the following replacement is to be done:
\begin{equation}
\ar^{(i)}(Q^2) \to \ar^{(i),{\rm fr}}(Q^2)
= \ar^{(i)}(Q^2 + M^2_{\rho})~~ (i=0,1,2),
\label{Intro:2}
\end{equation}
where the symbol $i+1$ marks the $i$th order of perturbation theory.

A second possibility is based on the idea proposed by Shirkov and
Solovtsov~\cite{ShiSo,SoShi} (see also recent
reviews~\cite{Cvetic} and references therein)
regarding the analyticity of the strong coupling constant in the
complex $Q^2$-plane in the form of the K\"all\'en-Lehmann spectral
representation. This approach leads effectively to additional
power $Q^2$-dependence for the DIS structure functions (see
Eq.~(\ref{B4})).

\subsection{Naive case}
\label{naive}
This modification is in a sense quite similar to the freezing procedure given above~(\ref{Intro:2}), i.e. the one- and two-loop coupling constants
$\alpha^{(0)}_s(Q^2)$ and $\alpha^{(1)}_s(Q^2)$ 
are to be replaced as follows:
\bea
\ar^{(0)}(Q^2) &\to& \ar^{(0),{\rm an}}(Q^2) =
        \ar^{(0)}(Q^2) - \frac{1}{\beta_0}\,
\frac{\Lambda^2_{(0)}}{Q^2 - \Lambda^2_{(0)}},
\label{an:LO} \\
\ar^{(1)}(Q^2) &\to& \ar^{(1),{\rm an}}(Q^2) =
        \ar^{(1)}(Q^2) - \frac{1}{2\beta_0}\,
\frac{\Lambda^2_{(1)}}{Q^2 - \Lambda^2_{(1)}}
+ \ldots\,,
\label{an:NLO}
\eea
where the ellipsis stands for cut terms which give negligible contributions
in our analysis.

At the one-loop level, the expression for the analytic coupling constant~(\ref{an:LO})
is very simple. However, at higher-loop levels, it has a rather cumbersome
structure (see a recent paper~\cite{Nesterenko} and discussions therein).
Therefore, it seems to be simpler to use some proper approximations so as to be able to carry out a numeric analysis.

Considering the study~\cite{SoShi}, in the NLO case the difference between
analytic and standard coupling constants can be represented in the form given
in~(\ref{an:NLO}), which is similar to the LO one with the additional coefficient equal to $1/2$.
Note that numerically this NLO term is quite analogous to the LO one, since
$\Lambda_{(0)} \ll \Lambda_{(1)}$ (see, for example,~\cite{Illarionov:2004nw}).

Following the logic expounded in~\cite{ShiTer}, where it was shown that
at the NNLO level the effective LO $\Lambda^{eff}_{(0)}$ can approximately be taken to be
\be
\Lambda^{eff}_{(0)}= (2\pi^2)^{\frac{-\beta_1}{2\beta_0^2}} \Lambda
\sim \frac{1}{2} \Lambda\,,~~~ \Lambda \equiv \Lambda_{(2)}
\ee
we can apply a simple analytic form in the NNLO as follows
\footnote{To have the poles in (\ref{an:NLO})
and~(\ref{an:NNLO}) exactly cancelled by those in the perturbative expansions
of QCD coupling (see, for example,~\cite{BKKP}), we keep
$Q^2-\Lambda^2_{(i)}$ ($i=0,1,2$) in the denominators of the
additional terms (i.e. in the last terms of (\ref{an:NLO}) and
(\ref{an:NNLO})) and, therefore, above LO we will have additional
terms coincinding with those in the LO case (with the
corresponding replacements $\Lambda_{0} \to \Lambda_{1}$ and
$\Lambda_{0} \to \Lambda$) multiplied by the additional factor
$1/(2i)$.} \be \ar(Q^2) \to \ar^{{\rm an}}(Q^2) =
        \ar(Q^2) - \frac{1}{4\beta_0}\,
\frac{\Lambda^2}{Q^2 - \Lambda^2}
+ \ldots\,, ~~~ \ar \equiv \ar^{(2)} \,.\label{an:NNLO}
\ee
Thus, we propose to use this last expression in the NNLO approximation.

In a sense, the replacement quoted in~(\ref{an:LO}),~(\ref{an:NLO}) and~(\ref{an:NNLO})
is a naive way of doing ``analytization''; we apply the latter procedure to 
the coupling constant itself without considering its actual function.
Nonetheless, this procedure has already
been successfully applied in~\cite{Zotov,CIKK09} for analyzing the DIS 
structure functions at small $x$ values. 
\footnote{The results obtained in \cite{CIKK09} show that naive analytization
as well as freezing of the coupling constant (\ref{Intro:2}) leads to the strong
improvement of the agreement with experimental data for the structure function $F_2$ and its 
slope at small $x$ values in the double asymptotic scaling regime 
(see \cite{Q2evo} and references therein).
Moreover, these results are very similar to the ones
 \cite{Illarionov:2004nw} obtained 
in the framework of the infrared renormalon model \cite{Beneke}
for HT corrections.}

With this motivation we would like to investigate its effect in the present study as well.
Here it will be referred to as a procedure of the {\it naive ``analytization''}.

\subsection{Transition to the canonical form}

To accomplish the procedure of ``analytization'' more accurately
\footnote{We will call this case an ordinary {\it analytic perturbation theory (APT)}.}, it is convenient for the moment $M_n^{NS}(Q^2)$ 
to be represented in the
following form (see \cite{Kotikov:2007ua} and discussions therein)
\be
M_n^{NS}(Q^2) = M_n^{NS}(Q^2_0)
\cdot {\left[\mu_n^{NS}(Q^2,Q^2_0)\right]}^{\frac{\gamma_{NS}^{(0)}(n)}{2\beta_0}}
\label{3.21a}
\ee
where the new ``moment'' $\mu_n^{NS}(Q^2)$ starts with $ \ar(Q^2)$ (see 
\cite{KKB}).

The procedure~(\ref{3.21a}) has already been used in~\cite{PKK,Vovk},
where the Grunberg's effective method~\cite{Grunberg} has been incorporated
into the analyses of DIS structure functions.

Now, the ``moment'' $\mu_n^{NS}(Q^2)$ has the form close to that
obtained for the sum rule, because it begins with 
the first power of $\ar(Q^2)$.
Consequently, the form gets closer to that
in the difference between the QCD sum rule and its Parton Model
value.
Following~\cite{ShiTer}, the analytical version of Eq.~(\ref{3.21a})
has the following form \cite{KKB}
\bea
\mu_{(0),n}^{NS}(Q^2,Q^2_0) &=& \frac{A_1^{(1)}(Q^2)}{A_1^{(1)}(Q^2_0)}\,,
\label{3.22a} \\
\mu_{(1),n}^{NS}(Q^2,Q^2_0) &=&
\frac{A_1^{(2)}(Q^2) + A_2^{(2)}(Q^2) \tilde{B}_{NS}^{(1)}(n)}
{A_1^{(2)}(Q^2_0) + A_2^{(2)}(Q^2_0) z^{(1)}_{NS}(n)}\,,  \nonumber \\
\mu_n^{NS}(Q^2,Q^2_0) &=&
\frac{A_1^{(3)}(Q^2) + A_2^{(3)}(Q^2) \tilde{B}_{NS}^{(1)}(n) +
A_3^{(3)}(Q^2) \tilde{B}^{(2)}_{NS}(n)}
{A_1^{(3)}(Q^2_0) + A_2^{(3)}(Q^2_0) z^{(1)}_{NS}(n) +
A_3^{(3)}(Q^2_0) z^{(2)}_{NS}(n)}\,,
\nonumber
\eea
where $A_m^{(i)}$ is the ``analytized'' $m$-th power of $i$-loop QCD 
coupling~\cite{ShiSo} and the coeffcients $\tilde{B}^{(i)}_{NS}(n)$ and
$z^{(i)}_{NS}(n)$ can be found in \cite{KKB}.

Thus, in the APT case the procedure features more complicated functions
rather than just the powers of some coupling constant.
In the one-loop case the  Euclidean functions of analytic perturbation theory (APT) $A_k^{(1)}(Q^2)$ are found to be~\cite{SoShi}
\bea
A^{(1)}_1(Q^2) &=& \frac{1}{\beta_0} \left[ \frac{1}{L_{(0)}} -
\frac{\Lambda^2_{(0)}}{Q^2-\Lambda^2_{(0)}}\right], 
~~ A^{(1)}_{k+1} ~=~ -\frac{1}{k\beta_0} \, \frac{d A^{(1)}_{k}}{dL_{(0)}},
\nonumber \\
A^{(1)}_2(Q^2) &=&
\frac{1}{\beta_0^2} \left[ \frac{1}{(L_{(0)})^2} -
\frac{\Lambda^2_{(0)}Q^2}{(Q^2-\Lambda^2_{(0)})^2}\right], \nonumber \\
A^{(1)}_3(Q^2) &=&
\frac{1}{\beta_0^3} \left[ \frac{1}{(L_{(0)})^3} -
\frac{\Lambda^2_{(0)}Q^2(Q^2+
\Lambda^2_{(0)})}{2(Q^2-\Lambda^2_{(0)})^3}\right]\,,
 \label{3.21e}
\end{eqnarray}
where $L_{(0)}=\ln(Q^2/(\Lambda_{(0)})^2)$, while
beyond the LO approximation~\cite{SoShi,ShiTer} the transform from standard
perturbation theory to the APT is slightly modified to assume the following form:
\bea
\ar^{(i)}(Q^2) \to
A^{(i+1)}_1(Q^2) &=& \ar^{(i)}(Q^2) -
\frac{1}{2i\beta_0}\,
\frac{\Lambda^2_{(i)}}{Q^2-\Lambda^2_{(i)}}\,, ~~~(i=1,2)
\nonumber \\
{\left(\ar^{(i)}(Q^2)\right)}^2 \to
A^{(i+1)}_2(Q^2) &=& \left(\ar^{(i)}(Q^2)\right)^2 -
\frac{1}{2i\beta^2_0}\,
\frac{\Lambda^2_{(i)}Q^2}{(Q^2-\Lambda^2_{(i)})^2}\,, \label{3.21l} \\
{\left(\ar^{(i)}(Q^2)\right)}^3 \to
A^{(i+1)}_3(Q^2) &=& \left(\ar^{(i)}(Q^2)\right)^3 -
\frac{1}{2i \beta^3_0}\, \frac{\Lambda^2_{(i)}Q^2(Q^2+
 \Lambda^2_{(i)})}{2(Q^2-\Lambda^2_{(i)})^3}\,,
\nonumber
\end{eqnarray}
where the NLO and NNLO results (for $i=1$ and $2$) are only some approximations.

To clear up with the meaning of all these formulas, we would like to note one more time
that in the expressions for $A^{(i)}_m(Q^2)$ the lower subscript stands for the power of the coupling
constant, while the upper one is related with the order of the approximation considered. Therefore, if the $\as$-expansion of some variable starts with the first power, as is the case at hand, then the power of the last term of the expansion coincides with the order of the approximation, i.e. for the last term upper and lower subscripts coincide.

\subsection{Fractional analytic perturbation theory}

Recently, in a series of papers~\cite{BMS}, the analytic continuation
(\ref{3.21e}) has been extended to the noninteger powers of LO
$L^{-1}_{(0)}$. In the one-loop case, this so-called fractional APT (FAPT) gives:
\be
\frac{1}{L^{\nu}_{(0)}}  \to \frac{1}{L^{\nu}_{(0)}} -
\frac{{\rm Li}_{1-\nu}(e^{-L_{(0)}})}{\Gamma(\nu)},
\label{f.1}
\ee
where
\be
{\rm Li}_{\nu}(z) = \sum_{m=1}^{\infty} \, \frac{z^m}{m^{\nu}}
\label{f.2}
\ee
is actually the polylogarithm function.~\footnote{In~\cite{BMS} it was called the Lerch transcendent function.}

It is clearly seen that by virtue of
\bea
{\rm Li}_{0}(z) = \frac{z}{1-z},\, {\rm Li}_{-N}(z) = \left(z\frac{\dd}{\dd z}\right)^N\frac{z}{1-z} ~=~ \left(z\frac{\dd}{\dd z}\right)^N\frac{1}{1-z}\,,
\label{f.3}
\eea
all the equations quoted in~(\ref{3.21e}) can be shown to be well reproduced.

Since the product of $\Gamma$-functions satisfy
$$
\Gamma{(\nu)}\Gamma{(1-\nu)} = \frac{\pi}{\sin(\pi \nu)}\,,
\label{f.4a}
$$
it is easy to obtain the following representation (see Appendix~A in 
\cite{KKB}):
\be
\frac{{\rm Li}_{1-\nu}(z)}{\Gamma(\nu)} ~=~
\frac{z\,\sin(\pi \nu)}{\pi} \,
\int\limits_{0}^1 \frac{{\mathrm d}\xi}{1-z\xi} \ln^{-\nu} \left(\frac{1}{\xi}\right),
\label{f.4}
\ee
that holds for $\Re{\mathrm e}(1-\nu) > 0$ and $\Re{\mathrm e}(\nu) > 0$ (in the case at hand these boil down to $0<\nu<1$).

Then, following the previous section let's recast Eq.(\ref{f.1}) in the following fashion:
\be
{\left(\ar^{(0)}(Q^2)\right)}^{\nu} \to
A^{(1)}_{\nu}(Q^2) ~=~ \left(\ar^{(0)}(Q^2)\right)^{\nu} -
\frac{{\rm Li}_{1-\nu}\left(\Lambda_{(0)}^2/Q^2\right)}{\beta_0^\nu\Gamma(\nu)}\,,
\label{f.5}
\ee
which reproduces the equations in~(\ref{3.21e}) with $\nu=1,2$ and $3$, in order.

Above LO, using quite the same arguments as in the previous subsection we have by analogy $(i=1,2)$:
\bea
{\left(\ar^{(i)}(Q^2)\right)}^{\nu} \to
A^{(i+1)}_{\nu}(Q^2) &=& \left(\ar^{(i)}(Q^2)\right)^{\nu} -
\frac{{\rm Li}_{1-\nu}\left(\Lambda_{(i)}^2/Q^2\right)}{2i\beta_0^\nu\Gamma(\nu)}\,,
\label{f.6}
\eea
which in turn reproduces a set of equations quoted in~(\ref{3.21l}) for $\nu=1,2$ and $3$, respectively.

Note that the Mellin moments
used in this case, contain\\
$\nu= \gamma_{NS}^{(0)}(n)/(2\beta_0)+i$ with $i=0$ in the LO, $i=0,1$ --- NLO, 
and $i=0,1,2$ --- NNLO approximations. The argument of the polylogarithm 
function $\Lambda^2_{(i)}/Q^2$ found in Eqs.~(\ref{f.5}) and~(\ref{f.6}) can 
be expressed through the strong coupling constant (see \cite{KKB}).
Therefore, the power $\nu$ lies within the range $0<\nu<4$, because
$0<\gamma_{NS}^{(0)}(n)/(2\beta_0)<2$ for $2<n<10$ used in the analyses.
The integral representation given in Eq.~(\ref{f.4}) is correct only for 
$0<\nu<1$, hence the need to extend it to higher values of $\nu$. 
Omitting details of this latter extension~\footnote{This is done in 
Appendix~A of \cite{KKB}, 
where $N=1,2,3$. Similar study can be found in the recent paper 
\cite{CveticKo}.} we have $(0<\delta <1)$
\be
\frac{{\rm Li}_{1-N-\delta}(z)}{\Gamma(N+\delta)} ~=~
{\left(z \frac{\mathrm{d}}{\mathrm{d}z}\right)}^N \, \left[
\frac{z}{\Gamma{(N+\delta)}\Gamma{(1-\delta)}} \,
 \int\limits_{0}^1 \frac{{\mathrm d}\xi}{1-z\xi} \ln^{-\delta} \left(\frac{1}{\xi}\right)  \right]
\,.
\label{f.7}
\ee

To cover an entire range $0<\nu<4$, we should consider $N=0, 1, 2$ and $3$. For
$N=0$, Eq.~(\ref{f.4}) with the corresponding replacement $\nu\to\delta$ can be used.
For $N>0$, it is straightforward to obtain
\bea
\left(z \frac{\mathrm{d}}{\mathrm{d}z}\right) \frac{z}{1-z\xi}
=  \frac{z}{(1-z\xi)^2}\,, & &
{\left(z \frac{\mathrm{d}}{\mathrm{d}z}\right)}^2 \frac{z}{1-z\xi}
=  \frac{z(1+z\xi)}{(1-z\xi)^3}\,, \nonumber \\
{\left(z \frac{\mathrm{d}}{\mathrm{d}z}\right)}^3 \frac{z}{1-z\xi}
&=&  \frac{z(1+4z\xi +z^2\xi^2)}{(1-z\xi)^4}
\label{f.8}
\eea
and make use of the following formulas valid for $0<\delta<1$:
\bea
\frac{{\rm Li}_{-\delta}(z)}{\Gamma(1+\delta)} &=&
\frac{z}{\Gamma{(1+\delta)}\Gamma{(1-\delta)}} \,
 \int\limits_{0}^1 \frac{{\mathrm d}\xi \ln^{-\delta}(1/\xi)}{(1-z\xi)^2}
, \label{f.9a} \\
\frac{{\rm Li}_{-1-\delta}(z)}{\Gamma(2+\delta)} &=&
\frac{z}{\Gamma{(2+\delta)}\Gamma{(1-\delta)}} \,
 \int\limits_{0}^1 \frac{{\mathrm d}\xi \ln^{-\delta}(1/\xi)}{(1-z\xi)^3}
\, \Bigl[1+z\xi\Bigr] , \label{f.9b} \\
\frac{{\rm Li}_{-2-\delta}(z)}{\Gamma(3+\delta)} &=&
\frac{z}{\Gamma{(3+\delta)}\Gamma{(1-\delta)}} \,
 \int\limits_{0}^1 \frac{{\mathrm d}\xi \ln^{-\delta}(1/\xi)}{(1-z\xi)^4}
\, \Bigl[1+4z\xi +z^2\xi^2\Bigr] , \label{f.9c}
\eea
i.e. Eqs.~(\ref{f.9a}),~(\ref{f.9b}) and~(\ref{f.9c}) can be used within the ranges
$1<\nu<2$, $2<\nu<3$ and $3<\nu<4$, respectively.


\section{ A fitting procedure }
\label{sec3}

A numeric procedure of fitting the data is described in the previous papers~\cite{BKKP,KK2001}. Here we just recall some aspects of the so-called polynomial expansion method.
The latter was first proposed in~\cite{Ynd} and further developed in~\cite{gon}. In these papers the method was based on the Bernstein polynomials and subsequently used to analyze data at NLO~\cite{KaKoYaF,KaKo} and NNLO level~\cite{SaYnd,KPS1}.
The Jacobi polynomials for that purpose were first proposed and then 
subsequently developed in~\cite{Barker,Kri} and used in~
\cite{PKK,KKPS1,KPS1,Vovk}

With the QCD expressions for the Mellin moments $M_n^{k}(Q^2)$
analytically calculated according to the formula in~(\ref{1.a}),
the SF $F_2^k(x,Q^2)$ is reconstructed by using the Jacobi polynomial expansion method:
$$
F_{2}^k(x,Q^2)=x^a(1-x)^b\sum_{n=0}^{N_{max}}\Theta_n ^{a,b}(x)\sum_{j=0}^{n}c_j^{(n)}(\alpha ,\beta )
M_{j+2}^k (Q^2)\,,
\label{2.1}
$$
where $\Theta_n^{a,b}$ are the Jacobi polynomials and $a,b$ are the parameters fitted. A condition put on the latter is the requirement of the error minimization while reconstructing the structure functions.

Since a twist expansion starts to be applicable only above $Q^2
\sim 1$ GeV$^2$ the cut $Q^2 \geq 1$ GeV$^2$ on the data is
applied throughout.

MINUIT program~\cite{MINUIT} is used to minimize two variables
$$
\chi^2 = \biggl|\frac{F_2^{exp} - F_2^{teor}}{\Delta
F_2^{exp}}\biggr|^2\,,
\qquad
\chi^2_{slope} = \biggl|\frac{D^{exp} - D^{teor}}{\Delta D^{exp}}\biggr|^2\,,
$$
where 
$D={\mathrm d}\ln F_2/{\mathrm d}\ln\ln Q^2$. The quality of the fits is characterized by $\chi^2/DOF$
for the structure function $F_2$. Analysis is also performed for the SF slope
$D$ that serves the purpose of checking the properties of fits (for more details see~\cite{BKKP}).

We use free normalizations of the data for different experiments.
For a reference set, the most stable deuterium BCDMS data at the
value of the beam initial energy $E_0=200$~GeV is used. With the
other datasets taken to be a reference one the variation in the
results is still negligible. In the case of the fixed
normalization for each and all datasets the fits tend to yield a
little bit worse $\chi^2$, just as in the previous studies.

\section{Results}

Since there is no gluons in the nonsinglet approximation the analysis is essentially
easier to conduct, with the cut imposed on the Bjorken variable $x\geq 0.25$
where gluon density is believed to be negligible.

Here we conduct separate and combined analyses of SLAC, BCDMS and NMC datasets for hydrogen and deuterium targets.
The cut on $x$ is imposed in a combination with those placed on the $y$ variable as follows:
\bea
& &y \geq 0.14 \,~~~\mbox{ for  }~~~ 0.3 < x \leq 0.4 \nonumber \\
& &y \geq 0.16 \,~~~\mbox{ for  }~~~ 0.4 < x \leq 0.5 \nonumber \\
& &y \geq 0.23 ~~~\mbox{ for  }~~~ 0.5 < x \leq 0.6 \nonumber \\
& &y \geq 0.24 ~~~\mbox{ for  }~~~ 0.6 < x \leq 0.7 \nonumber \\
& &y \geq 0.25 ~~~\mbox{ for  }~~~ 0.7 < x \leq 0.8\,, \nonumber
\eea
which are meant to cut out those points with large systematic errors. Thus, upon imposing the cuts
a complete dataset  consists of 327 points in the case of hydrogen target and 288 --- deuterium one. The starting point of the QCD evolution is taken to be $Q^2_0=90$~GeV$^2$.
This $Q^2_0$ value is close to the average values of $Q^2$ spanning the corresponding data.  From earlier studies~\cite{BKKP,KK2001} it follows that it is enough to take the maximal value of the number of moments to be accounted for $N_{max} =8$~\cite{Kri}; also note that the cut $0.25 \leq x \leq 0.8$ is imposed everywhere.

To reduce the number of parameters, we perform two groups of fits.
The first one is dealt within the variable-flavor-number-scheme
(VFNS)~\cite{BKKP} and the $H_2$ and $D_2$  experimental datasets analyzed simultaneously.
The results for the second group are obtained within the fixed-flavor-number-scheme (FFNS) with an active number of flavors $n_f=4$ and the $H_2$ and $D_2$ experimental datasets considered separately.

\subsection{VFNS case}

All the results obtained within our reference VFNS~\cite{BKKP} and in the cases of ``naive'', APT, and ``frozen'' modifications of $\as$ are gathered in a set of tables separately for each order of perturbation theory approximation and displayed in Figs.~1--3 separately for 
all these three cases.
\vskip0.5cm
{\bf Table 1.} {\sl LO values of the twist-four term $\tilde h_4(x)$
(with statistic errors given) obtained in the analysis
of the combined $H_2 + D_2$ dataset within the VFNS
and with various modifications of $\as$ }
\begin{center}
\scriptsize
\begin{tabular}{|c||c|c|c|c|c|c|c|c|}
\hline
$x$ & \multicolumn{2}{c|}{Naive analyt. $\as$ } & \multicolumn{2}{c|}{APT-inspired
$\as$ } & \multicolumn{2}{c|}{Frozen $\as$}  & \multicolumn{2}{c|}{Standard $\as$} \\
\hline \hline
0.275&\multicolumn{2}{c|}{-0.121 $\pm$ 0.008}&\multicolumn{2}{c|}{-0.123 $\pm$ 0.008}&\multicolumn{2}{c|}{-0.204 $\pm$ 0.011} &\multicolumn{2}{c|}{-0.271 $\pm$0.012}\\
0.35 &\multicolumn{2}{c|}{-0.055 $\pm$ 0.007}&\multicolumn{2}{c|}{-0.055 $\pm$ 0.007}&\multicolumn{2}{c|}{-0.167 $\pm$ 0.017} &\multicolumn{2}{c|}{-0.257 $\pm$0.017}\\
0.45 &\multicolumn{2}{c|}{0.119 $\pm$ 0.012}&\multicolumn{2}{c|}{0.119 $\pm$ 0.012}&\multicolumn{2}{c|}{-0.021 $\pm$ 0.031} &\multicolumn{2}{c|}{-0.144 $\pm$0.030}\\
0.55 &\multicolumn{2}{c|}{0.422 $\pm$ 0.022}&\multicolumn{2}{c|}{0.422 $\pm$ 0.023}&\multicolumn{2}{c|}{0.211 $\pm$ 0.053} &\multicolumn{2}{c|}{0.051 $\pm$0.049}\\
0.65 &\multicolumn{2}{c|}{0.870 $\pm$ 0.060}&\multicolumn{2}{c|}{0.866 $\pm$ 0.059}&\multicolumn{2}{c|}{0.558 $\pm$ 0.095} &\multicolumn{2}{c|}{0.364 $\pm$0.088}\\
0.75 &\multicolumn{2}{c|}{1.322 $\pm$ 0.117}&\multicolumn{2}{c|}{1.336 $\pm$ 0.112}&\multicolumn{2}{c|}{0.917 $\pm$ 0.152} &\multicolumn{2}{c|}{0.709 $\pm$0.138}\\
\hline
$\chi^2/DOF$ &\multicolumn{2}{c|}{0.93}&\multicolumn{2}{c|}{0.93}&\multicolumn{2}{c|}{0.91} &\multicolumn{2}{c|}{0.94}\\
$\chi^2_{slope}/DOF$ &\multicolumn{2}{c|}{2.30}&\multicolumn{2}{c|}{2.35}&\multicolumn{2}{c|}{2.15} &\multicolumn{2}{c|}{2.60}\\
$\asMZ $ &\multicolumn{2}{c|}{0.1474}&\multicolumn{2}{c|}{0.1474}&\multicolumn{2}{c|}{0.1409} &\multicolumn{2}{c|}{0.1400}\\
\hline
\end{tabular}
\end{center}

{\bf Table 2.} {\sl NLO values of the twist-four term
$\tilde h_4(x)$ (with statistic errors given)
obtained in the analysis of the combined $H_2 + D_2$ dataset
within the VFNS and with various modifications of $\as$ }
\begin{center}
\scriptsize
\begin{tabular}{|c||c|c|c|c|c|c|c|c|}
\hline
$x$& \multicolumn{2}{c|}{Naive analyt. $\as$} & \multicolumn{2}{c|}{APT-inspired
$\as$} & \multicolumn{2}{c|}{Frozen $\as$} & \multicolumn{2}{c|}{Standard $\as$} \\
\hline \hline
0.275&\multicolumn{2}{c|}{-0.127 $\pm$ 0.009}&\multicolumn{2}{c|}{-0.129 $\pm$ 0.007}&\multicolumn{2}{c|}{-0.183 $\pm$ 0.008}&\multicolumn{2}{c|}{-0.229 $\pm$ 0.010}\\
0.35 &\multicolumn{2}{c|}{-0.098 $\pm$ 0.007}&\multicolumn{2}{c|}{-0.024 $\pm$ 0.009}&\multicolumn{2}{c|}{-0.149 $\pm$ 0.010}&\multicolumn{2}{c|}{-0.218 $\pm$ 0.016}\\
0.45 &\multicolumn{2}{c|}{0.014 $\pm$ 0.012}&\multicolumn{2}{c|}{0.187 $\pm$ 0.013}&\multicolumn{2}{c|}{ 0.010 $\pm$ 0.019}&\multicolumn{2}{c|}{-0.084 $\pm$ 0.030}\\
0.55 &\multicolumn{2}{c|}{0.172 $\pm$ 0.024}&\multicolumn{2}{c|}{0.506 $\pm$ 0.019}&\multicolumn{2}{c|}{0.215 $\pm$ 0.033}&\multicolumn{2}{c|}{0.098 $\pm$ 0.052}\\
0.65 &\multicolumn{2}{c|}{0.339 $\pm$ 0.057}&\multicolumn{2}{c|}{0.910 $\pm$ 0.045}&\multicolumn{2}{c|}{0.476 $\pm$ 0.065}&\multicolumn{2}{c|}{0.356 $\pm$ 0.093}\\
0.75 &\multicolumn{2}{c|}{0.478 $\pm$ 0.107}&\multicolumn{2}{c|}{1.230 $\pm$ 0.090}&\multicolumn{2}{c|}{0.757 $\pm$ 0.108}&\multicolumn{2}{c|}{0.648 $\pm$ 0.145}\\
\hline
$\chi^2/DOF$ &\multicolumn{2}{c|}{0.85}&\multicolumn{2}{c|}{0.84}&\multicolumn{2}{c|}{0.97}&\multicolumn{2}{c|}{1.02}\\
$\chi^2_{slope}/DOF$ &\multicolumn{2}{c|}{0.82}&\multicolumn{2}{c|}{0.78}&\multicolumn{2}{c|}{0.87} &\multicolumn{2}{c|}{1.20}\\
$\asMZ $ &\multicolumn{2}{c|}{0.1275}&\multicolumn{2}{c|}{0.1224}&\multicolumn{2}{c|}{0.1169}&\multicolumn{2}{c|}{0.1152}\\
\hline
\end{tabular}
\end{center}
 \vskip0.5cm

{\bf Table 3.} {\sl NNLO values of the twist-four term $\tilde h_4(x)$
(with statistic errors given) obtained in the analysis
of the combined $H_2 + D_2$ dataset within the VFNS and with various
modifications of $\as$ }
\begin{center}
\scriptsize
\begin{tabular}{|c||c|c|c|c|c|c|c|c|}
\hline
$x$ & \multicolumn{2}{c|}{Naive analyt. $\as$} & \multicolumn{2}{c|}{APT-inspired
$\as$} & \multicolumn{2}{c|}{Frozen $\as$} & \multicolumn{2}{c|}{Standard $\as$}  \\
\hline \hline
0.275&\multicolumn{2}{c|}{-0.171 $\pm$ 0.006}&\multicolumn{2}{c|}{-0.196 $\pm$ 0.008}&\multicolumn{2}{c|}{-0.149 $\pm$ 0.006}&\multicolumn{2}{c|}{-0.173 $\pm$ 0.017}\\
0.35 &\multicolumn{2}{c|}{-0.160 $\pm$ 0.008}&\multicolumn{2}{c|}{-0.152 $\pm$ 0.012}&\multicolumn{2}{c|}{-0.129 $\pm$ 0.013}&\multicolumn{2}{c|}{-0.094 $\pm$ 0.020}\\
0.45 &\multicolumn{2}{c|}{-0.044 $\pm$ 0.018}&\multicolumn{2}{c|}{0.030 $\pm$ 0.022}&\multicolumn{2}{c|}{-0.007 $\pm$ 0.031}&\multicolumn{2}{c|}{-0.110 $\pm$ 0.015}\\
0.55 &\multicolumn{2}{c|}{0.085 $\pm$ 0.033}&\multicolumn{2}{c|}{0.269 $\pm$ 0.038}&\multicolumn{2}{c|}{0.116 $\pm$ 0.062}&\multicolumn{2}{c|}{-0.086 $\pm$ 0.033}\\
0.65 &\multicolumn{2}{c|}{0.221 $\pm$ 0.065}&\multicolumn{2}{c|}{0.551 $\pm$ 0.074}&\multicolumn{2}{c|}{0.218 $\pm$ 0.115}&\multicolumn{2}{c|}{0.085 $\pm$ 0.083}\\
0.75 &\multicolumn{2}{c|}{0.304 $\pm$ 0.100}&\multicolumn{2}{c|}{0.782 $\pm$ 0.116}&\multicolumn{2}{c|}{0.258 $\pm$ 0.169}&\multicolumn{2}{c|}{0.158 $\pm$ 0.105}\\
\hline
$\chi^2/DOF$ &\multicolumn{2}{c|}{0.98}&\multicolumn{2}{c|}{1.02}&\multicolumn{2}{c|}{0.97}&\multicolumn{2}{c|}{0.92}\\
$\chi^2_{slope}/DOF$ &\multicolumn{2}{c|}{1.22}&\multicolumn{2}{c|}{1.17}&\multicolumn{2}{c|}{1.02} &\multicolumn{2}{c|}{1.83}\\
$\asMZ $ &\multicolumn{2}{c|}{0.1151}&\multicolumn{2}{c|}{0.1125}&\multicolumn{2}{c|}{0.1163}&\multicolumn{2}{c|}{0.1159}\\
\hline
\end{tabular}
\end{center}
\vskip0.5cm

From Tables~1--3 it is seen that in all the cases considered
$\chi^2/DOF$ shows
good agreement between the experimental data and
theoretical predictions for the SF Mellin moments. In LO and NLO
cases, the analytic and ``frozen'' modifications lead to some 
additional improvement of fits, namely, $\chi^2/DOF$
in these cases is less than that in the standard case.

Since the quantity $\chi^2_{slope}/DOF$ is inherently linked with the pQCD aspects to be observed in the data it in this respect is very informative, for it strongly varies from one approximation to the other thus indicating if there are any effects incompatible with the $Q^2$ dependence in each $x$-bin assumed. It is seen that it, much like the $\chi^2/DOF$  quantity, demonstrates similar tendency: in all the cases considered, the analytic and ``frozen''  modifications lead to improvement of fits. 
Certain improvement of the quality of fits is observed at the NLO level.
In this approximation even the standard $\as$ case leads to reasonable agreement for the slopes; furthermore, for all the infrared modifications it is seen that $\chi^2_{slope}/DOF <1$. In the NNLO approximation the numbers for the slope in the case with a standard $\as$ are not as good but the infrared modifications (especially the ``frozen'' one) lead to better
agreement between theoretical and experimental results for the former quantity.
It is difficult to pin down a reason for such a deterioration 
of this agreement (in the scheme with a standard $\as$) when NNLO corrections are added. Because this effect is absent in the FFNS case (see the following subsection), we suppose
that the exact equation for the coupling constant, used here,
is somehow
in inconsistency with the NNLO expression for the heavy quark thresholds, which is
in fact based on certain expansions of the coupling constant 
at the threshold crossing points (see, for example,~\cite{BKKP}).
We plan to study this fine effect elsewhere.

The QCD coupling decreases from LO through NNLO,
which is in perfect agreement with other studies
(see~\cite{Kotikov:2007ua} and references therein); it is seen that 
the frozen modification gives the results closest to those for the standard $\as$.
In the analytic cases the values of $\asMZ$ are higher
in the first two orders of perturbation theory, 
\footnote{This observation is consistent with earlier studies
(see~\cite{Shi2001} and references therein).}
with the maxima in the difference $\Delta\asMZ$ being $0.0123$ and $0.0072$ 
for the naive analytic and APT cases, respectively, observed at NLO. 
Thus, the differences are substantially greater than
the values of the total experimental error
$\Delta^{total}\asMZ=0.0022$, which was obtained in~\cite{BKKP} by
combining statistical and systematic errors in quadrature.

At the NNLO the APT procedure leads to lower $\asMZ$, though in the naive case 
the $\asMZ$ value is closer to those obtained for the frozen and standard versions.
Nevertheless, all the NNLO $\asMZ$ values, except for the APT case, 
are in good agreement within statistical
errors, which were found to be $\Delta^{stat}\asMZ=0.0007$ in our
previous studies~\cite{BKKP}. The APT-inspired QCD coupling
constant is compatible with the rest within the total experimental
error.

From Tables~1--3 and, particularly, Figs.~1--3 it is seen that in
the cases of analytic and ``frozen'' coupling constants the
twist-four corrections are larger compared to those in the case of
a standard perturbative coupling constant, thus confirming the
results obtained in~\cite{ShiTer}.
For example, at $x\sim 0.75$ the higher-twist corrections (HTCs)
in the standard case are about twice as less that those obtained
in the cases of analytic modifications in all orders considered.
In the ``frozen'' case, the HTC values are compatible with those in
the standard case within statistical errors.~\footnote{For the
results to be discernible, Figs.~1--7 contain 
statistical  errors for the infrared modifications only. The
magnitude of the errors is on par in the standard case and is not
shown.}

The difference in the values of HT parameter for the naive analytic
and frozen variants becomes moderate at the NLO level unlike the APT case. 
For the infrared modifications the HT terms are large 
and in the frozen and APT cases are compatible with the LO ones.
A partial explanation of this effect can be found in Appendix~A. The 
``analytization'' of the coupling constant generates additional
power-like contributions with the opposite sign as compared
to the twist-four corrections at large $x$ and, thus, increases
effectively the values of HTCs. The analysis in Appendix~A
is given at the LO level. Above the latter 
the corresponding results, in accordance with Eqs.~(\ref{an:LO}),~(\ref{an:NLO}) and~(\ref{an:NNLO}), can be estimated by replacing
$\Lambda_{(0)} \to \Lambda_{(1)}/2$ in NLO
and $\Lambda_{(0)} \to \Lambda/4$ in NNLO in front of the term $\ln(1-x)$
in~(\ref{B4}).

In the NNLO approximation the situation changes drastically. HT terms
for all the cases, except for the APT case, are comparable with each other.
HTCs in the latter case are still higher but they are also strongly suppressed
compared to HTCs in the LO and NLO cases. To some extent, this
can be explained by replacing $\Lambda_{(0)} \to \Lambda/4$ in front of $\ln(1-x)$
in~(\ref{B4}), which decreases the influence of the infrared modifications.

Thus, at the NNLO level the twist-four corrections appear
to be small for all the cases considered above. 

The values for parameters in the parameterizations of the parton
distributions 
for the cases corresponding
to different coupling constant modifications are given in Table~4 of \cite{KKB}.

\subsection{FFNS case}

For comparison let's present the values of PDF parameters obtained
within the FFNS ($n_f=4$) in the analyses of the hydrogen and deuterium data for
the versions of the strong coupling constant discussed above, 
with the addition for the case of FAPT-inspired modification of $\as$.
Here we have no threshold transitions for QCD coupling and PDF
Mellin moments and, hence, are able to consider the hydrogen and
deuterium datasets separately. 

The values for parameters in the parameterizations of the parton
distributions 
for the cases corresponding
to different coupling constant modifications are given in Table~5 of \cite{KKB}.


In order to be able to assess the difference among the cases with frozen, 
FAPT and standard versions for the strong coupling constant let's 
present the tables with the HT values obtained within FFNS ($n_f=4$)
for the hydrogen data (results obtained for the deuterium data can be found 
in \cite{KKB}).

\vskip0.5cm
{\bf Table 4.} {\sl LO values of the twist-four term obtained within the 
FFNS ($n_f=4$) and with various modifications of $\as$. Only statistical
errors are given.
}
\begin{center}
\scriptsize
\begin{tabular}{|c||c||c||c|}
\hline
& FAPT-inspired $\as$ & Frozen $\as$  & Standard $\as$ \\
\hline
$x$ & $\tilde h_4(x)$ & $\tilde h_4(x)$ & $\tilde h_4(x)$ \\
\hline \hline
0.275&-0.221$\pm$0.011&
-0.183 $\pm$ 0.012 &-0.235$\pm$0.012
\\
0.35 &-0.187$\pm$0.014&
-0.160 $\pm$ 0.018 &-0.232$\pm$0.021
\\
0.45 & 0.002$\pm$0.023&
-0.023 $\pm$ 0.037 &-0.130$\pm$0.038
\\
0.55 & 0.332$\pm$0.034& 
0.189 $\pm$ 0.065  & 0.049$\pm$0.065
\\
0.65 & 1.001$\pm$0.063& 
0.610 $\pm$ 0.117 & 0.455$\pm$0.111
\\
0.75 & 2.031$\pm$0.131& 
1.177 $\pm$ 0.207 & 1.003$\pm$0.182
\\
\hline
$\chi^2/DOF$ & 0.98 &
0.94 & 0.98   \\
$\chi^2_{slope}/DOF$ & 1.47 & 1.25 & 1.58  \\
$\asMZ $ &0.1394&
0.1387& 0.1376 
\\
\hline
\end{tabular}
\end{center}

\vskip0.5cm
{\bf Table 5.} {\sl NLO values of the twist-four term obtained within the FFNS ($n_f=4$) and with various modifications of $\as$.Only statistical
errors are given.}
\begin{center}
\scriptsize
\begin{tabular}{|c||c||c||c|}
\hline
& FAPT-inspired $\as$ & Frozen $\as$ & Standard $\as$   \\
\hline
$x$ & $\tilde h_4(x)$ & $\tilde h_4(x)$ & $\tilde h_4(x)$ for $H_2$ \\
\hline \hline
0.275&-0.205$\pm$0.012&
-0.154 $\pm$ 0.011 &-0.200$\pm$0.012\\
0.35 &-0.194$\pm$0.017& 
-0.147 $\pm$ 0.016 &-0.219$\pm$0.019\\
0.45 &-0.067$\pm$0.030& 
-0.058 $\pm$ 0.037 &-0.169$\pm$0.041\\
0.55 & 0.160$\pm$0.044& 
0.069 $\pm$ 0.067  &-0.078$\pm$0.072\\
0.65 & 0.635$\pm$0.071& 
0.317 $\pm$ 0.118& 0.153$\pm$0.123\\
0.75 & 1.512$\pm$0.130& 
0.723 $\pm$ 0.204 & 0.534$\pm$0.205\\
\hline
$\chi^2/DOF$ & 0.91 &
0.89 & 0.92  \\
$\chi^2_{slope}/DOF$ & 1.08 & 0.92 & 1.23  \\
$\asMZ $ &0.1220
0.1200& 0.1192  \\
\hline
\end{tabular}
\end{center}

\vskip0.5cm
{\bf Table 6.} {\sl NNLO values of the twist-four term obtained within the FFNS ($n_f=4$) and with various modifications of $\as$.Only statistical
errors are given.}
\begin{center}
\scriptsize
\begin{tabular}{|c||c||c||c|}
\hline
& FAPT-inspired $\as$ & Frozen $\as$ & Standard
$\as$  \\
\hline
$x$ & $\tilde h_4(x)$ & $\tilde h_4(x)$& $\tilde h_4(x)$ \\
\hline \hline
0.275&-0.168$\pm$0.010&
-0.119 $\pm$ 0.011 & -0.158 $\pm$ 0.020 \\
0.35 &-0.181$\pm$0.015&
-0.112 $\pm$ 0.010 & -0.166 $\pm$ 0.021 \\
0.45 &-0.130$\pm$0.037&
-0.049 $\pm$ 0.018 &-0.156 $\pm$ 0.036  \\
0.55 &-0.056$\pm$0.065&
0.007 $\pm$ 0.031 & -0.157 $\pm$ 0.079 \\
0.65 & 0.126$\pm$0.111& 
0.106 $\pm$ 0.058  & -0.061 $\pm$ 0.127 \\
0.75 & 0.419$\pm$0.178& 
0.240 $\pm$ 0.116 &  0.049 $\pm$ 0.209\\
\hline
$\chi^2/DOF$ & 0.89 & 
0.87 & 0.89  \\
$\chi^2_{slope}/DOF$ & 0.97 & 0.68 & 0.97  \\
$\asMZ $ &0.1183&
0.1180& 0.1176 \\
\hline
\end{tabular}
\end{center}

Just like in the previous subsection it is seen from Tables~4--6
that in all the cases at hand, we have good agreement 
between the experimental data and theoretical predictions. 
Once again, in all the cases considered, excluding LO FAPT, 
the analytic and frozen modifications lead to slight improvement
of fits, that is 
$\chi^2/DOF$ and $\chi^2_{slope}/DOF$ are found to be smaller in these cases 
than those for the standard $\as$, and the QCD coupling constant
decreases when we move from LO through NNLO.

Also note that contrary to what was observed in the previous subsection 
here the quantity $\chi^2_{slope}/DOF$ steadily decreases when we proceed step by step from LO to NLO and then to NNLO level. Respectively, here the smaller values of $\chi^2_{slope}/DOF$ are observed for the ``frozen'' case as well.

The values of coupling constants are very similar, especially for the ``frozen''
and standard versions. In the analytic cases, the central values of
$\asMZ$ are little higher, mostly in the first two orders of perturbation 
theory (see also~\cite{Shi2001}).


In the NNLO, 
all the $\asMZ$ values, except for the
analytic one, are in good agreement within statistical
errors.
As earlier, the analytic QCD coupling is in agreement
with the rest within the total experimental error.

Similarly to the previous subsection we note in
Figs.~4,~5,~6~
appreciable difference between HTCs obtained in the standard and analytic
cases in the first two orders of perturbation theory. 
The difference in the cases of a ``frozen'' and standard $\as$ is
not as large; the values of HT terms are in agreement within
statistical errors (see Fig.~7).

In the NNLO, the situation is over again changed considerably. In
all the cases of infrared modifications considered, the HTCs are
small. Similarly to the previous subsection they
are not compatible with zero at $x \sim 0.75$, while the HT terms
in standard QCD are compatible with zero and, at the same time, in
agreement with all the cases considered within  statistical
errors.

Then, at NNLO we can see agreement between standard QCD and its
infrared modifications for QCD coupling $\asMZ$, as well as for
the respective HTCs (as a rule) within statistical errors.

\section{Conclusions}

The pattern of separating perturbative QCD and HT corrections may be
different in different orders of perturbation theory, as well as in some
resummations based on several first orders, and for certain
modifications of the strong coupling constant as well.
In the present paper, we have shown the results of our recent analysis 
\cite{KKB}, where we studied the consequences of the infrared modifications
of the QCD coupling constant --- the so-called ``frozen'' and ``analytized''
versions. In the last case 
three different options were considered:
\begin{itemize}
\item{a simple modification of the strong coupling constant~\cite{ShiSo}
without rearrangement of a perturbation series;}
\item{an application of the ordinary analytic perturbation theory 
(see~\cite{ShiTer}) to the ``moments'' $\mu_n^{NS}(Q^2)$ given in~(\ref{3.21a});}
\item{impact of the fractional analytic perturbation theory~\cite{BMS}
applied directly to the Mellin moments $M_n^{NS}(Q^2)$.}
\end{itemize}
To test all these modifications,
the Jacobi polynomial expansion method developed in~\cite{Barker,
Kri} was used to perform analysis of $Q^2$-evolution of the
DIS structure function $F_2$ by fitting all existing to date
reliable fixed-target experimental data that satisfy the cut $x
\geq 0.25$. To the best of our knowledge, the study \cite{KKB}
is the first
application of the FAPT results to the fits of the DIS structure
functions.

The main conclusions are as follows. In the first two orders of
perturbation theory, the coupling constant $\asMZ$ rises for
all versions of the analytic perturbation theory.
This observation is in complete agreement with
recent studies in~\cite{Shi2001}.

An increase in the
central values of $\asMZ$ is smaller in the NNLO approximation, much like
in the case of the ``frozen'' version of $\as$. Nevertheless, in all
the cases the $\asMZ$  values are in agreement mostly within total
experimental errors. For the NNLO case, the results are as a rule
compatible between each other within statistical errors.
Also, we note that within statistical errors only (in our case 
$\Delta^{stat}\asMZ=0.0007$~\cite{BKKP}) all the NNLO results 
for $\as(M_Z^2)$ in the FFNS case (see Table~6) are 
too compatible with the world average value for the
coupling constant presented in the review ~\cite{Breview}
\footnote{It should be mentioned that this analysis was carried out over
the data coming from the various experiments and in different orders of
perturbation theory, i.e. from NLO up to N$^{3}$LO.}:
$$ 
\as(M_Z^2) = 0.1184 \pm 0.0007\,.
$$

It is also observed that there is some rise of the twist-four corrections in the first two
orders of perturbation theory, particularly for all versions of
the analytic one. This observation is in complete agreement with
recent studies~\cite{ShiTer} of the Bjorken sum rules,
where it was shown that there is a reduction of higher HT terms, 
starting with the twist-six ones. Unfortunately, in \cite{KKB} we were
not able to study
the twist-four and twist-six corrections simultaneously since their contributions 
are strongly correlated.

In the NNLO, all the cases of infrared modifications of the QCD
coupling feature nonzero although rather small
twist-four corrections. 
Whereas in the case of the standard QCD approach, the HT terms are
close to zero at large values of the Bjorken variable $x$. 
However, the NNLO HTCs for all the cases considered are compatible between each other
within statistical errors.

In principle, the main difference between the cases with a standard QCD coupling constant
and its analytic and ``frozen'' modifications is in the strong suppression of
the higher twist corrections in NLO and NNLO orders of
perturbation theory, respectively.

What is interesting to look for further in the study 
is the consideration of the combined
nonsinglet and singlet analyses using the DIS experimental data
within an entire $x$ region, as well as an application of certain
resummation-like Grunberg effective charge methods~\cite{Grunberg}
(as was done in~\cite{Vovk} in the NLO approximation) and the
``frozen''~\cite{frozen}~\footnote{There are a lot of ``frozen''
versions of the strong coupling constant (see, for example, the
list of references in \cite{Zotov}).} and analytic~\cite{SoShi}
versions of the strong coupling constant (see~\cite{CIKK09,Zotov,ShiTer} 
for recent studies in this direction). The effect of N$^3$LO corrections
would also seem to be important to account for in the subsequent investigations,
as well as an extension of the FAPT model results for $\as$ to the VFNS case.

\section{Acknowledgments}
The work was supported by the RFBR grant No.10-02-01259-a.
We are grateful to S.V.~Mikhailov, D.B.~Stamenov and O.V.~Teryaev for useful discussions.

\section{Appendix A}
\label{App:B}
\def\theequation{A\arabic{equation}}
\setcounter{equation}{0}

Consider the large $x$ asymptotic~\cite{Gross} of SF $F_2(x,Q^2)$ (for simplicity
we restrict ourselves to the LO approximation)
\be
F_2(x,Q^2) \sim (1-x)^{b(a^{(0)}_s(Q^2))},\label{B1}
\ee
where
$$
b\left(a^{(0)}_s(Q^2)\right) = b_0 - \tilde{d}
\ln (a^{(0)}_s(Q^2)), ~~ \tilde{d} = \frac{16}{3\beta_0}\,,
$$
and $b_0$ is some constant which can be obtained from the quark counting rules~\cite{schot}.

The $Q^2$-dependent part in the r.h.s. of~(\ref{B1}) can be represented as follows:
$$
(1-x)^{- \tilde{d}\ln (a^{(0)}_s(Q^2))} =
{\left[a^{(0)}_s(Q^2)\right]}^{- \tilde{d}\ln (1-x)}\,.
$$

Performing the ``analytization'' procedure and using~(\ref{an:LO}), we have
\footnote{This analysis is carried out at rather large $Q^2$ values used
in the paper. For low $Q^2$, consideration can be found, for example, in
\cite{CveKo}.
}
\bea
&&{\left(\ar^{(0)}(Q^2) - \frac{1}{\beta_0}\,
\frac{\Lambda^2_{(0)}}{Q^2 - \Lambda^2_{(0)}}\right)}^{- \tilde{d}\ln (1-x)}
\nonumber \\
&&\approx {\left[a^{(0)}_s(Q^2)\right]}^{- \tilde{d}\ln (1-x)} \left[1+
\frac{\tilde{d}\ln (1-x)}{\beta_0 \ar^{(0)}(Q^2)}
\frac{\Lambda^2_{(0)}}{Q^2 - \Lambda^2_{(0)}} \right]\,.
\label{B4}
\eea
From this expression it follows that the power corrections, which were generated
by the ``analytization'' of the coupling constant, have the opposite sign as compared
to the twist-four corrections at large $x$ and, moreover, demonstrate
a different asymptotic behavior.
These corrections (taken with the additional sign ``$-$'') increase like
$\ln(1-x)$ while the twist-four corrections behave like $1/(1-x)$ (see~\cite{Yndu}
and discussions therein). This difference in sign between these
corrections leads to the larger twist-four corrections in the present
analysis than those given in~\cite{BKKP}.


\newpage

\begin{figure}
\unitlength=1mm
\vskip -1.5cm
\begin{picture}(0,100)
\put(0,-5){\psfig{file=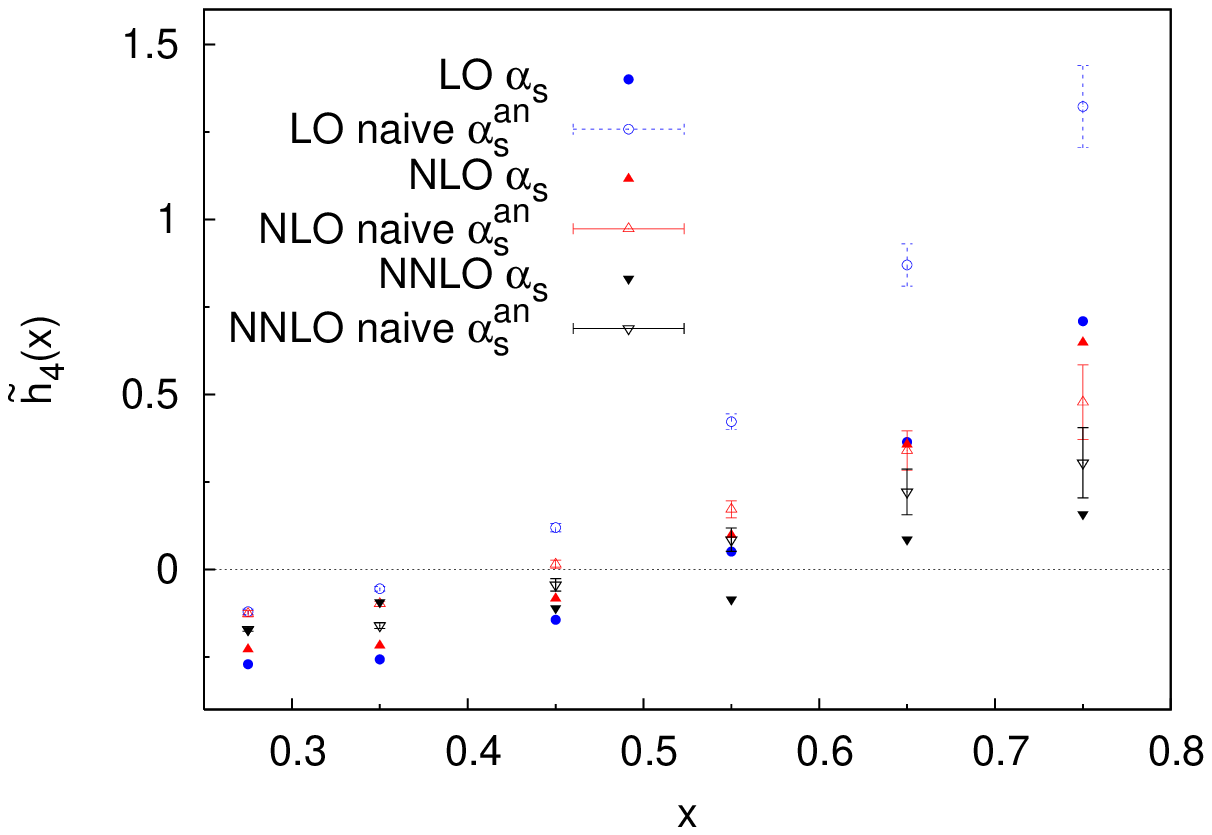,width=150mm,height=95mm}}
\end{picture}
\vskip 0.2cm
\caption{ Comparison of the HTC parameter $\tilde h_4(x)$ obtained in
LO, NLO and NNLO for hydrogen data (the bars stand for statistical errors) between our reference VFNS with a standard perturbative $\as$~\cite{BKKP} and that with naive analytic $\as$.}
\end{figure}

\begin{figure}
\unitlength=1mm
\vskip -1.5cm
\begin{picture}(0,100)
 \put(0,-5){%
  \psfig{file=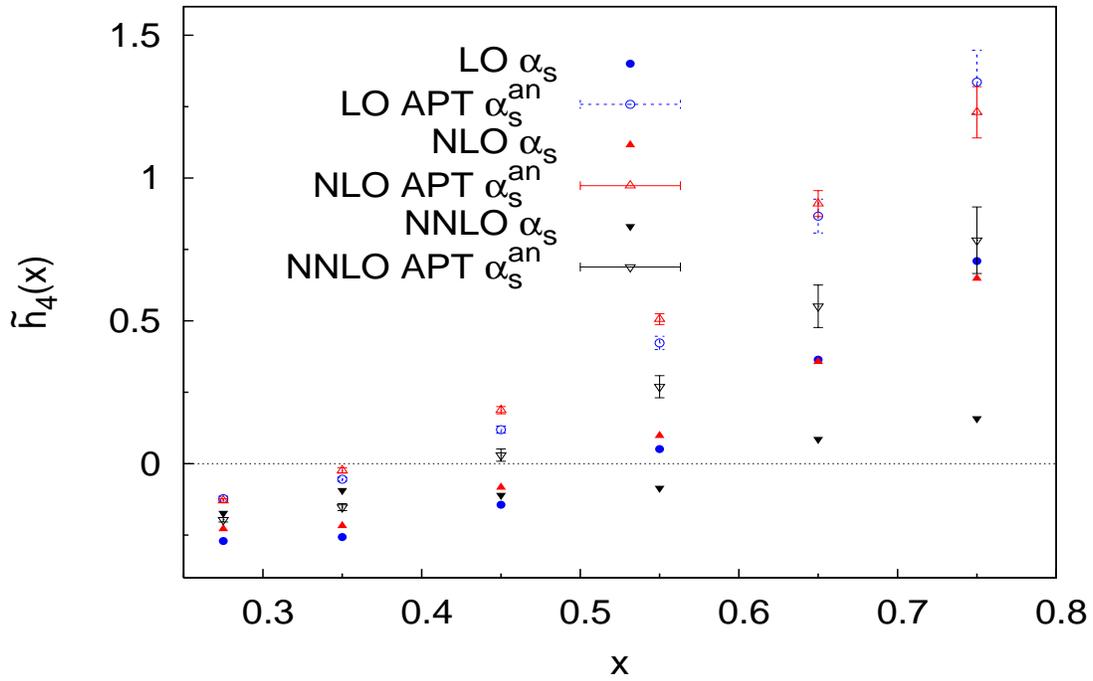,width=150mm,height=95mm}}
\end{picture}
\vskip 0.2cm
\caption{Comparison of the HTC parameter $\tilde h_4(x)$ obtained in
LO, NLO and NNLO for hydrogen data between a VFNS and that with APT-inspired $\as$.}
\end{figure}

\begin{figure}
\unitlength=1mm
\vskip -1.5cm
\begin{picture}(0,100)
 \put(0,-5){\psfig{file=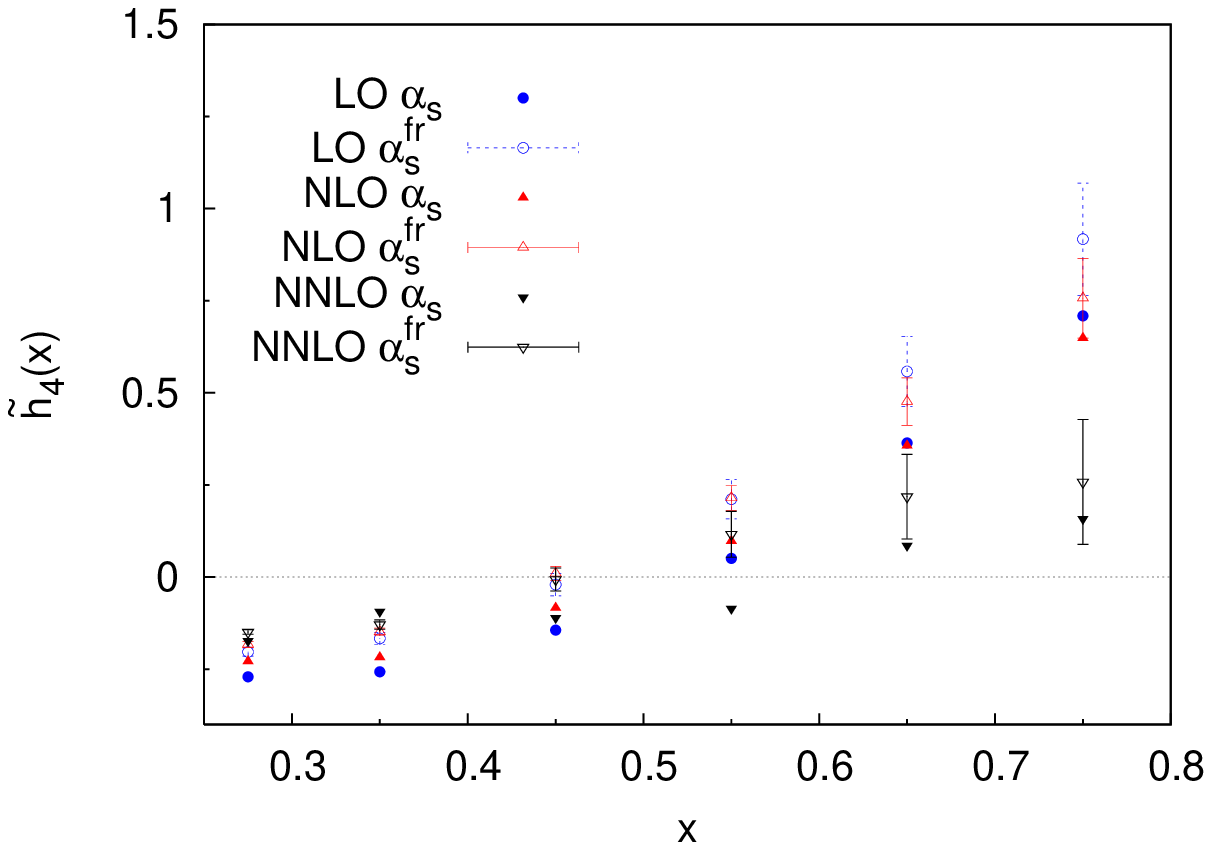,width=150mm,height=95mm}}
\end{picture}
\vskip 0.2cm
\caption{ Comparison of the HTC parameter $\tilde h_4(x)$ obtained at
LO, NLO and NNLO for hydrogen data within a VFNS between the cases with a standard perturbative and ``frozen''  $\as$.}
\end{figure}

\begin{figure}
\unitlength=1mm
\vskip -1.5cm
\begin{picture}(0,100)
\put(0,-5){\psfig{file=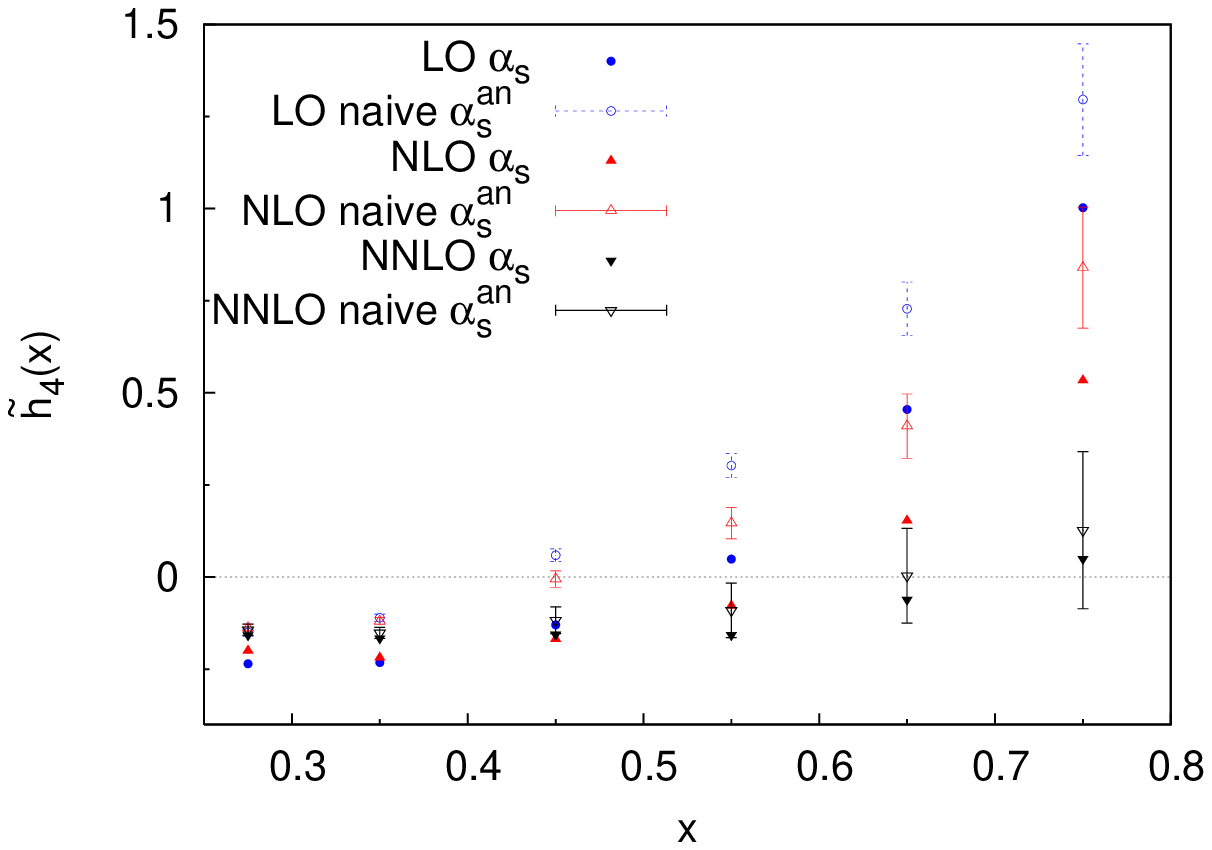,width=150mm,height=95mm}}
\end{picture}
\vskip 0.2cm
\caption{ Comparison of the HTC parameter $\tilde h_4(x)$ obtained in
LO, NLO and NNLO for hydrogen data within a FFNS ($n_f=4$) between the cases with a standard perturbative and naively analytized $\as$.} 
\end{figure}


\begin{figure}
\unitlength=1mm
\vskip -1.5cm
\begin{picture}(0,100)
 \put(0,-5){%
  \psfig{file=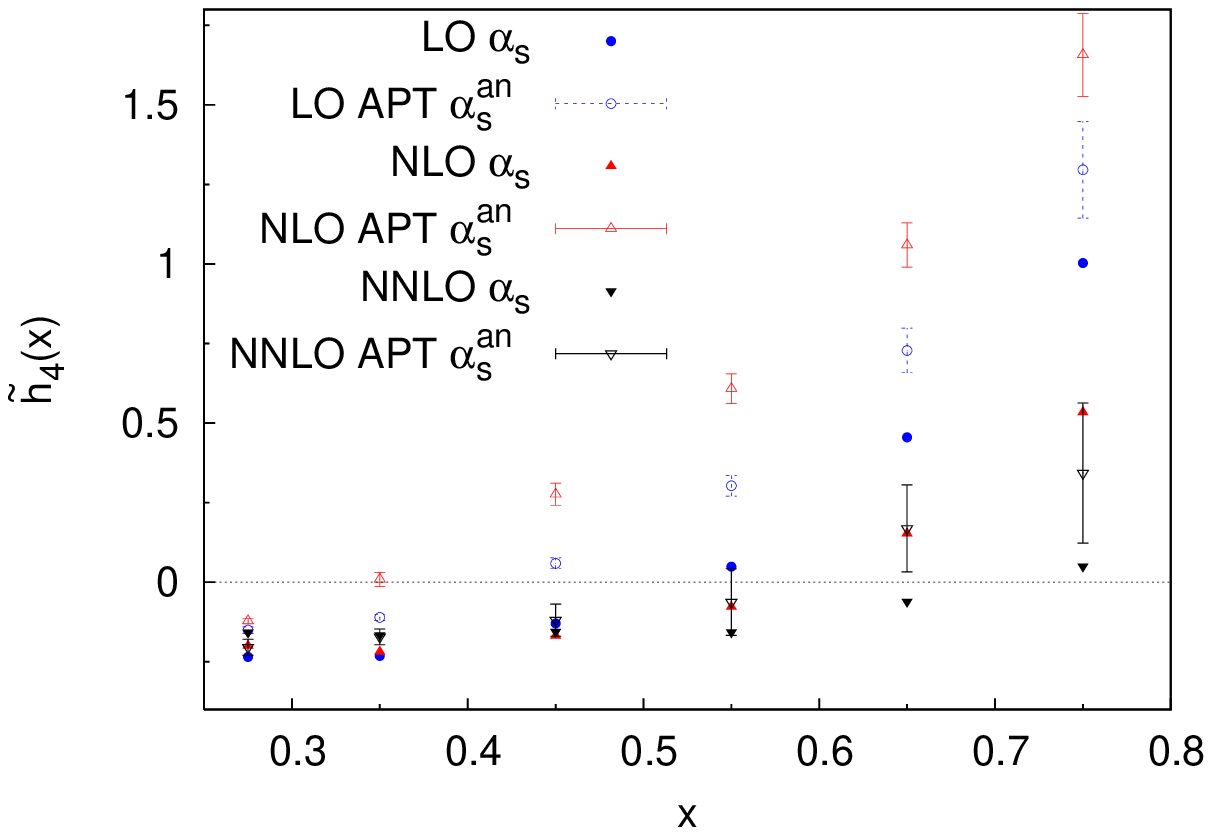,width=150mm,height=95mm}  }
\end{picture}
\vskip 0.2cm
\caption{ Comparison of the HTC parameter $\tilde h_4(x)$ obtained in
LO, NLO and NNLO for hydrogen data within a FFNS ($n_f=4$) between the cases with a standard perturbative and APT-inspired $\as$.}
\end{figure}



\begin{figure}
\unitlength=1mm
\vskip -1.5cm
\begin{picture}(0,100)
\put(0,-5){\psfig{file=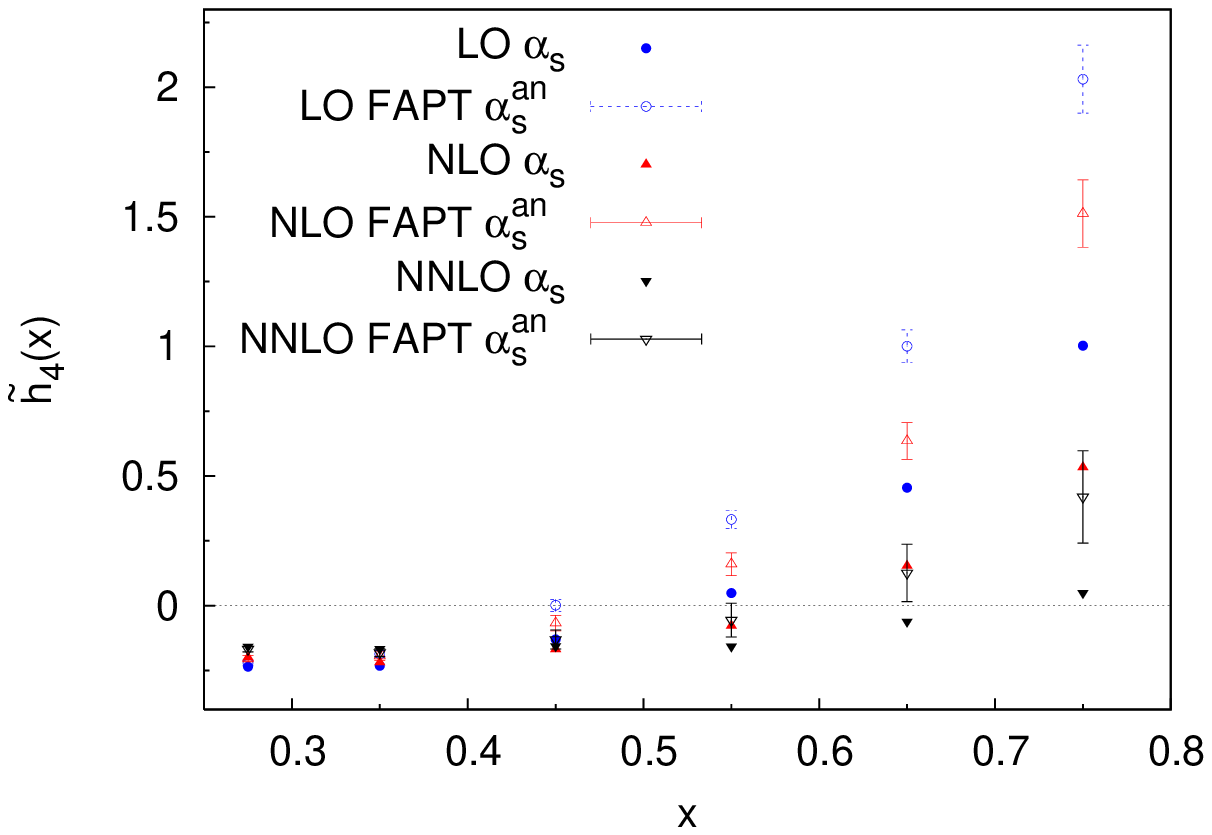,width=150mm,height=95mm}}
\end{picture}
\vskip 0.2cm
\caption{ Comparison of the HTC parameter $\tilde h_4(x)$ obtained at
LO, NLO and NNLO for hydrogen data within a FFNS ($n_f=4$)  between the cases with a standard perturbative and FAPT-inspired $\as$.}
\end{figure}

\begin{figure}
\unitlength=1mm
\vskip -1.5cm
\begin{picture}(0,100)
 \put(0,-5){%
  \psfig{file=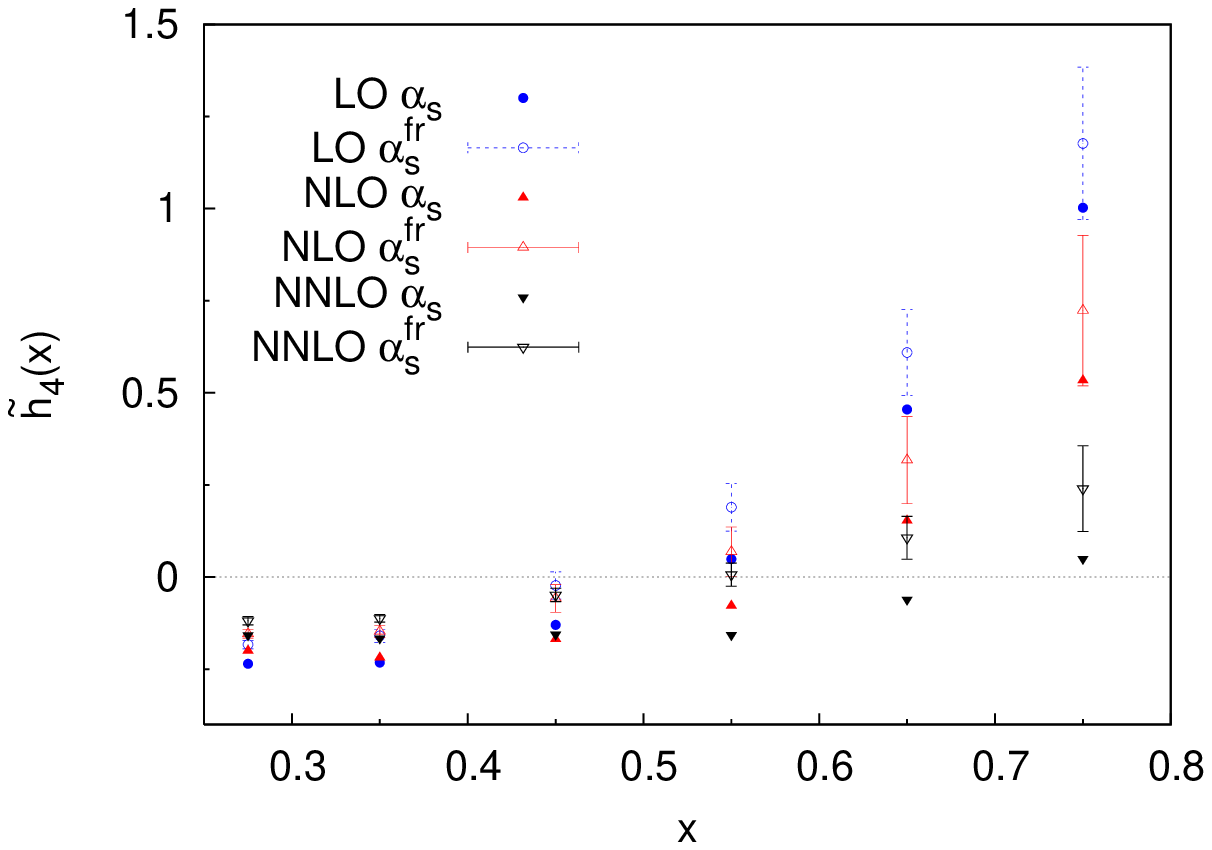,width=150mm,height=95mm}  }
\end{picture}
\vskip 0.2cm
\caption{ Comparison of the HTC parameter $\tilde h_4(x)$ obtained at
LO, NLO and NNLO for hydrogen data within a FFNS ($n_f=4$)  between the cases with a standard perturbative and ``frozen'' $\as$.}
\end{figure}


\end{document}